\documentclass[
twocolumn,
prx,
10pt,
amsmath,
amssymb,
nofootinbib,
showpacs,
superscriptaddress,
floatfix,
longbibliography
]{revtex4-1}

\usepackage{graphicx}
\usepackage{color}
\usepackage[usenames,dvipsnames]{xcolor}
\usepackage[colorlinks=true,
linkcolor=blue,
citecolor=blue,
linktoc=page]{hyperref}
\usepackage{multirow}
\usepackage{float}
\usepackage{flushend}
\usepackage{balance}
\usepackage[varg]{txfonts}
\usepackage{fancyhdr}
\usepackage{braket}
\usepackage{mathtools}
\usepackage{lipsum}

\begin{document}

\title{Linear Ascending Metrological Algorithm}

\author{M.\,R.~Perelshtein}
    \altaffiliation{These authors contributed equally: M.\,R.~Perelshtein and N.\,S.~Kirsanov}
    \affiliation{Terra Quantum AG, St.\,Gallerstrasse 16A, 9400 Rorschach, Switzerland}
    \affiliation{Moscow Institute of Physics and Technology, 141700, Institutskii Per.\ 9, Dolgoprudny, Moscow Distr., Russian Federation}
    \affiliation{QTF Centre of Excellence, Department of Applied Physics, Aalto University School of Science, P.O. Box 15100, FI-00076 AALTO, Finland}

\author{N.\,S.~Kirsanov}
    \altaffiliation{These authors contributed equally: M.\,R.~Perelshtein and N.\,S.~Kirsanov}
    \affiliation{Terra Quantum AG, St.\,Gallerstrasse 16A, 9400 Rorschach, Switzerland}
    \affiliation{Moscow Institute of Physics and Technology, 141700, Institutskii Per.\ 9, Dolgoprudny, Moscow Distr., Russian Federation}
    \affiliation{QTF Centre of Excellence, Department of Applied Physics, Aalto University School of Science, P.O. Box 15100, FI-00076 AALTO, Finland}
    \affiliation{Consortium for Advanced Science and Engineering (CASE) University of Chicago, 5801 S Ellis Ave, Chicago, IL 60637, USA}

\author{V.\,V.~Zemlyanov}
    \affiliation{Terra Quantum AG, St.\,Gallerstrasse 16A, 9400 Rorschach, Switzerland}
    \affiliation{Moscow Institute of Physics and Technology, 141700, Institutskii Per.\ 9, Dolgoprudny, Moscow Distr., Russian Federation}

\author{A.\,V.~Lebedev}
    \affiliation{Moscow Institute of Physics and Technology, 141700, Institutskii Per.\ 9, Dolgoprudny, Moscow Distr., Russian Federation}

\author{G.~Blatter}
    \affiliation{Theoretische Physik, Wolfgang-Pauli-Strasse 27, ETH Z\"{u}rich, CH-8093 Z\"{u}rich, Switzerland}

\author{V.\,M.~Vinokur}
    \affiliation{Consortium for Advanced Science and Engineering (CASE) University of Chicago, 5801 S Ellis Ave, Chicago, IL 60637, USA}
    \affiliation{Materials Science Division, Argonne National Laboratory, 9700 S. Cass Avenue, Argonne, Illinois 60637, USA}

\author{G.\,B.~Lesovik}
    \affiliation{Terra Quantum AG, St.\,Gallerstrasse 16A, 9400 Rorschach, Switzerland}
    \affiliation{Moscow Institute of Physics and Technology, 141700, Institutskii Per.\ 9, Dolgoprudny, Moscow Distr., Russian Federation}

\date{\today}

\begin{abstract}
The ubiquitous presence of shot noise sets a fundamental limit to
the measurement precision in classical metrology.  Recent advances in quantum
devices and novel quantum algorithms utilizing interference effects are
opening new routes for overcoming the detrimental noise
tyranny.  However, further progress is limited by the restricted capability
of existing algorithms to account for the decoherence pervading experimental
implementations.  Here, adopting a systematic approach to the
evaluation of effectiveness of metrological procedures, we devise the Linear Ascending Metrological Algorithm
(LAMA) which offers a remarkable increase of precision in the demanding
situation where a decohering quantum system is used to measure a continuously
distributed variable. We introduce our protocol in the context of magnetic
field measurements, assuming superconducting transmon devices as sensors
operated in a qudit mode. Our findings demonstrate a quantum
metrological procedure capable to mitigate detrimental dephasing and
relaxation effects.
\end{abstract}

\maketitle

\section{Introduction}
The shot noise sets a fundamental limit to
the measurement precision, which is often referred to as ``noise tyranny."  Quantum algorithms utilizing interference effects offer new opportunities to overcome it\,\cite{giovannetti2011, lesovik2010, suslov, degen1, degen2, sekatski,zemlyanov}.  However, existing algorithms cannot account for the decoherence controlled experimental conditions\,\cite{danilin}.  Our work develops a systematic approach in evaluation of the effectiveness of metrological procedures and seeking an
efficient solutions. To that end, we devise the Linear Ascending Metrological Algorithm
(LAMA) applicable and guaranteeing an enhanced  precision in the case where a decoherent quantum system is used to measure a continuously
distributed variable. We introduce our protocol in the context of magnetic
field measurements, assuming superconducting transmon devices as sensors
operated in a qudit mode \cite{shlyakhov}. 

Phase estimation protocols first appeared in abstract quantum algorithms,
where they served in estimating the phases of a unitary operator's
eigenvectors, and soon found practical applications in quantum
metrology \cite{giovannetti2004, vaidman, giedke, said, higgins, waldherr,
budker}.  Among many implementations, the Kitaev- \cite{kitaev, cleve} and
Fourier-transform- \cite{NC} algorithms, in combination with superconducting
transmon circuits \cite{koch} utilized as sensors, proved most efficient in
magnetometry.  The basic concept of a magnetic-field sensor based on a spin
interacting with the field has evolved into experimentally realizable devices
based on charge- and flux-qudits \cite{danilin, il'ichev, bal, wang}.  However,
the standard quantum-metrological protocols concede their optimal precision in
the most relevant situation  where the field is continuously distributed and
the quantum sensor suffers from decoherence.  Here, we devise a simple and
practical protocol for qudits -- LAMA -- allowing to appreciably enhance the
efficiency of the measurement over the standard Fourier-transform- and Kitaev
algorithms.  Distinct from the latter, our protocol benefits from a maximum
average spin component perpendicular to the field and takes advantage from a
linear step-wise increase of the Ramsey delay-time interval.  Throughout our
analysis, the operation of the different metrological algorithms will be
addressed in the context of their qutrit \cite{abdumalikov, kumar} (base-3)
transmon realization, as, unlike the example of a qubit (base-2) realization,
this allows us to demonstrate the new algorithm's full potential.

\section{General phase-sensitive protocol}
We begin with a description of the general base-$d$ phase-sensitive
metrological procedure employing a sequential strategy, with each step
following the Preparation-Exposure-Readout (PER) logic. The procedure is aimed
at the measurement of a constant magnetic field $H$.  We work with the
computational basis states $\ket{0}$, $\ket{1}$, \dots $\ket{d-1}$
corresponding to different magnetic components $\mathfrak{M}^d_Z$ with respect
to the field direction (Z-axis): for instance, in the qutrit case, the basis
vectors $\ket{0}$, $\ket{1}$, and $\ket{2}$ correspond, respectively, to
$\mathfrak{M}^3_Z$$=$$-\mu,0,+\mu$, where $\mu$ denotes the magnetic moment of
the artificial atom \cite{danilin}, which serves as a coupling constant and which is known a priori.  The
$i$th step of the general procedure involves a Ramsey interference with delay
time $t_i$ and is described as follows:
\begin{itemize}
    \item[\textbf{P}~] The qudit is prepared in a defined initial state
    $\ket{\psi^0_{(i)}}$; this is experimentally realized by applying a
    suitable rf-pulse to the qudit ground-state \cite{shlyakhov}.
    
    \item[\textbf{E}~] The qudit interacts with the external magnetic field
    $H$ during a time $t_i$: $\ket{\psi^{0}_{(i)}}$$\to$$
    \ket{\psi^{t_i}_{(i)}}$.  The field changes the phases inside the state
    vector such that the basis state $\ket{k}$ ($k=\{0,\,1,\,\dots,\,d-1\}$)
    transforms into $e^{i k \omega t_i}\ket{k}$, where $\omega=\mu H/{\hbar}$
    is the \textit{reduced} magnetic field; we omit a common phase factor, see
    Appendix \ref{transmon} for details on the transmon phase accumulation dynamics.
    
    \item[\textbf{R}~] The qudit is subjected to a readout operation (unitary
    evolution) $\hat{U}^r_{(i)}$ generated by another
    rf-pulse \cite{shlyakhov}.  The information about the field value is
    subsequently extracted through the final state's single-shot projective
    measurement in the computational basis.
\end{itemize}

\begin{figure}[t]
  \noindent\centering{
    \includegraphics[width=86mm]{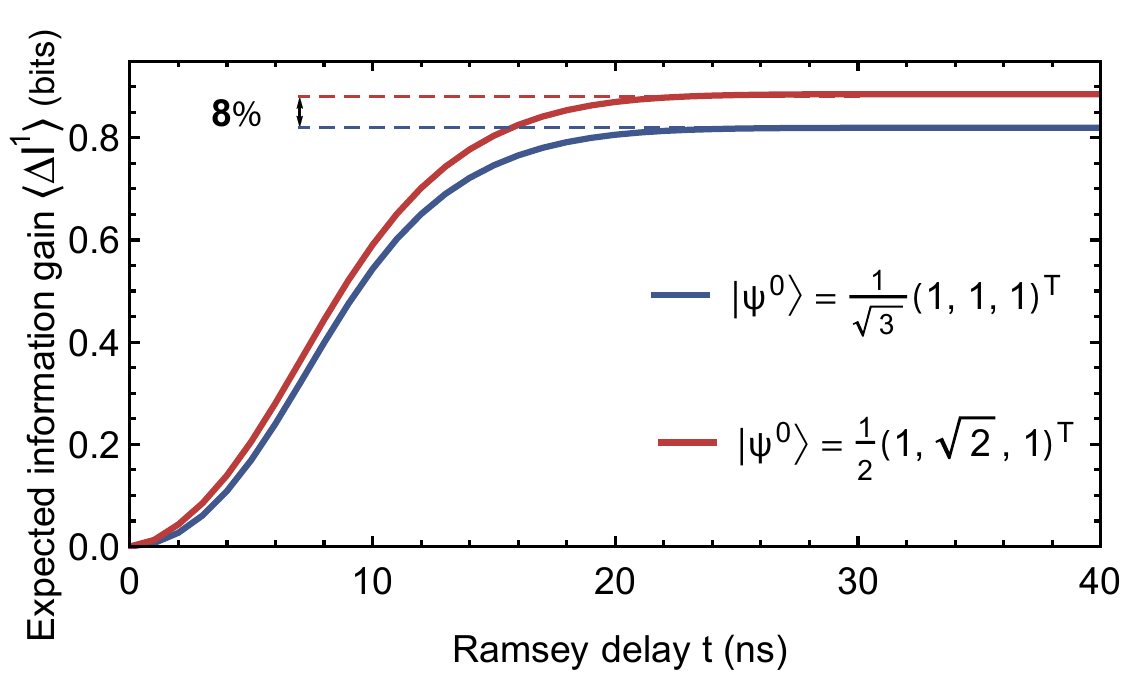}
    }
    \caption{Information gain for different initial states. Expected
    information gain during the first measurement step assuming a continuously
    distributed magnetic field: blue and red lines correspond to the
    situations where the qutrit is initially prepared in the balanced state
    (as in the standard Fourier procedure) and in the state
    $(1,\,\sqrt{2},\,1)^T/2\in \mathcal{H}^3_{XY}$, respectively.  The initial
    field distribution is $\mathcal{P}^{0}\big(\omega\,|\,\o\big) =
    N(0,\,\sigma^2)$ with $\sigma=2\pi/$($90$\,ns).  The saturation time is
    defined by the initial distribution of the magnetic field, {\color{black} $T_s\sim
    1/\sigma$}; the plateau-levels of the blue and red curves are,
    respectively, $(5/3$$-$$\ln{3})/\ln{2}$$ \approx$$ 0.82$\,bit and
    $2((\ln{2})^{-1}$$-$$1)$$ \approx$$ 0.88$\,bit. } \label{new_v_old}
\end{figure}
The probability of finding the qudit in the state $\ket{\xi_i}$
($\xi_i=\{0,\,1,\,\dots,\,d-1\}$) at the end of the $i$th step is
\begin{equation}
\label{outcome}
   P(\xi_i\,|\,\omega, t_i, \textbf{s}_i)
   =|\bra{\xi_i}\hat{U}^r_{(i)}\ket{\psi^{t_i}_{(i)}}|^2,
\end{equation}
where $\textbf{s}_i$ is the array of parameters determining the initial state
and the readout operation, see Appendix \ref{optimization}.  Throughout the metrological
procedure, or 'learning' process, our knowledge about the field is reflected
in the probability distribution $\mathcal{P}^n \big(\omega\,|\,\{\xi_i,t_i,
\textbf{s}_i\}_{i=1}^n \big)$, where $n$ indicates the number of conducted PER
steps.  This distribution is updated in accord with Bayes theorem via the
recurrence
\begin{multline}
\label{recurrent}
   \mathcal{P}^n \big(\omega\,|\,\{\xi_i,t_i,\textbf{s}_i\}_{i=1}^n\big) 
   \\=\mathcal{P}^{n-1}\big(\omega\,|\,\{\xi_i,t_i,\textbf{s}_i\}_{i=1}^{n-1}\big)\, 
   P(\xi_n\,|\,\omega, t_n, \textbf{s}_n)\,\mathcal{N}_n,
\end{multline}
where $\mathcal{N}_n$ is a normalization factor; for ease of presentation (see
Appendix \ref{oscillations_sec}), we will assume that the initial field
distribution is Gaussian with zero mean, $\mathcal{P}^0 \big(\omega\,
|\,\o\big)= N(0,\,\sigma^2)$. As shown below, the initial field uncertainty
$\delta\omega_0 = \sigma$ determines the difficulty of the further refinement: the
smaller $\delta\omega_0$ is, the harder it is to achieve better precision.

\section{Efficient procedure}
In an efficient metrological procedure, a measurement step strongly reduces
the uncertainty with regard to the possible field values. This uncertainty is
reflected in the Shannon entropy associated with $\mathcal{P}^n$,
\begin{multline}
\label{H}
   S^n\big(\{\xi_i,t_i,\textbf{s}_i\}_{i=1}^{n}\big) \\
   = -\int \mathcal{P}^n \big(\omega^\prime|\{\xi_i,t_i,\textbf{s}_i\}_{i=1}^n\big) 
   \ln{\mathcal{P}^n \big(\omega^\prime|\{\xi_i,t_i,\textbf{s}_i\}_{i=1}^n\big)
   d\omega^\prime}.
\end{multline}
The decrease $\Delta I^{n}=S^{n-1}-S^{n}$ in entropy then provides us with the
information gain in the $n$th measurement step.  The optimal procedure
yielding the maximum information gain in the $(n+1)$st step requires the best
choice of the parameters $t_{n+1}$ and $\textbf{s}_{n+1}$.  Since the values
of parameters are required before the step is executed, the optimization has
to be done by maximizing an estimate of the prospective information gain.
This estimate is taken as the information gain after the $(n+1)$st step
averaged over all possible outcomes $\tilde\xi_{n+1}$ of this step.
\begin{multline}
    \label{AvGain}
    \langle  \Delta I^{n+1}\big(\{\xi_i,t_i,\textbf{s}_i\}_{i=1}^{n}, 
    \{\tilde\xi_{n+1}, t, \textbf{s}\} \big) \rangle\\ 
    =  S^n\big(\{\xi_i,t_i,\textbf{s}_i\}_{i=1}^{n}\big) 
    -\langle  S^{n+1}\big(\{\xi_i,t_i,\textbf{s}_i\}_{i=1}^{n}, 
    \{\tilde\xi_{n+1}, t, \textbf{s}\} \big) \rangle,
\end{multline}
where
\begin{multline}
\label{AvEnt}
   \langle  S^{n+1}\big(\{\xi_i,t_i,\textbf{s}_i\}_{i=1}^{n}, 
   \{\tilde\xi_{n+1}, t, \textbf{s}\} \big) \rangle \\
   = \sum_{\tilde\xi_{n+1}=0}^{d-1} \int S^{n+1}\big(\{\xi_i,t_i,\textbf{s}_i\}_{i=1}^{n}, 
   \{\tilde\xi_{n+1}, t, \textbf{s}\} \big)\\
   \times P(\tilde\xi_{n+1}\,|\,\omega^\prime, t, \textbf{s}) \,\mathcal{P}^n 
   \big(\omega^\prime\,|\,\{\xi_i,t_i,\textbf{s}_i\}_{i=1}^n\big)\,d\omega^\prime.
\end{multline}
{\color{black}The optimal choice of parameters $\{t,\,\textbf{s}\} = \{t_{n+1},\,\textbf{s}_{n+1}\}$ is dictated by the condition of maximizing the average gain $\langle\Delta
I^{n+1}\big(\{\xi_i,t_i,\textbf{s}_i\}_{i=1}^{n}, \{\tilde\xi_{n+1},
t,\,\textbf{s}\} \big) \rangle$.
In other words, the analytical expression for the average information gain serves as a prognosis for the upcoming information gain, and its maximization yields the most beneficial choice of parameters.}

In order to find the optimal choice for $\textbf{s}_1$ defining the
preparation and readout gates in the first step, we maximize the expected
information gain $\langle \Delta I^{1} \big(\tilde\xi_{1}, t, \textbf{s} \big)
\rangle$ for every possible value of the Ramsey delay time $t$.  Focusing on
the case of a {\it continuously} distributed field \textcolor{black}{(i.e., $\omega$ can take any value from a certain interval $[\omega_\mathrm{min},\,\omega_\mathrm{max}]$)} measured with a qutrit
sensor, a numerical analysis (see Appendix \ref{optimization}) shows that, intriguingly, for any
$t$ the optimal initial state preparation requires a maximum {\color{black}modulus of the
spin projection into the XY-plane, perpendicular to the field vector, $\langle J_{XY} \rangle=\sqrt{\langle \hat{J}_{X} \rangle^2+\langle \hat{J}_{Y} \rangle^2}$ (here $\hat{J}_{X}$ and $\hat{J}_{Y}$ are, respectively, X- and Y-component spin operators)}:
let $\mathcal{H}^d_{XY}$ be the subspace of such vectors for the qudit
(base-d) system. For a qutrit, any vector $\ket{\phi}\in \mathcal{H}^3_{XY}$
can be written in the form $\ket{\phi} = \frac{e^{i\alpha}}{2} (e^{i\beta},
\,\sqrt{2}, \,e^{-i\beta})^T$, where $\alpha$ and $\beta$ are real numbers.  A
convenient choice is the initial state $\ket{\psi^0_{(1)}} = (1, \,\sqrt{2},
\,1)^T/2\in \mathcal{H}^3_{XY}$, which is the eigenstate of the spin operator
$\hat{J}_X$.  For comparison, in the standard Fourier-based procedure (see Ref.\
 \cite{shlyakhov} and Appendix \ref{fourier}) as optimized for the measurement of a {\it
discretely} distributed field \textcolor{black}{(i.e., $\omega$ can only take $M$ values $\{\omega_\mathrm{min},
\,\omega_\mathrm{min}+\Delta \omega,\, \omega_\mathrm{min}+2\Delta
\omega,\,\dots,\,\omega_\mathrm{max}\}$ with $\Delta \omega =
(\omega_\mathrm{max}-\omega_\mathrm{min})/M$)} the qutrit is initially prepared in the
balanced state \cite{suslov} $(1,\,1,\,1)^T/\sqrt{3} \notin
\mathcal{H}^3_{XY}$.  Physically, the larger spin component perpendicular to
the field {\color{black}($\langle J_{XY} \rangle = 1$ in the optimal case versus $2\sqrt{2}/3\approx 0.94$ in the case of the balanced state)} ensures a better sensitivity;
note, that in the qubit case, the balanced initial state $(1,\,1)^T/\sqrt{2}$
has already the largest perpendicular component and thus cannot be further
optimised. Turning to the readout gate, our analysis shows that, for any $t$,
the optimum is always achieved with the Fourier transform gate
    \begin{equation}
    \label{F}
        \hat{F}_3=\frac{1}{\sqrt{3}}\begin{pmatrix} 
                1 & 1 & 1 \\
                1 & e^{4\pi i/3} & e^{2\pi i/3} \\
                1 & e^{2\pi i/3} & e^{4\pi i/3}
        \end{pmatrix};
    \end{equation}
we remind that
for the discretely distributed field the Fourier transform readout operation
is also optimal \cite{suslov}. 

The difference between the standard Fourier-based procedure and the \textcolor{black}{described new optimal procedure (with the initial qutrit state $\ket{\psi^0}=(1,\,\sqrt{2},\,1)^T/2\in \mathcal{H}^3_{XY}$)} in the continuous situation is illustrated in Fig.\,\ref{new_v_old},
where we plot the expected information gain $\langle \Delta
I^{1}\big(\tilde\xi_{1}, t, \textbf{s} \big) \rangle$ in the first step as a
function of the Ramsey delay time $t$ with $\textbf{s}$ corresponding to
different initial states and the standard Fourier transform readout operation.
In both cases the information gain saturates at the time 
{\color{black} $T_s \sim 1/\delta\omega_0$} which is defined by the initial field uncertainty
$\delta\omega_0=\sigma$; a small $\delta\omega_0$ increases $T_s$, what makes
it increasingly harder to further improve the precision, particularly in the
presence of decoherence.
For the described optimal procedure, the curve reaches a
plateau that is higher by 8\%, with an information gain of
$2((\ln{2})^{-1}$$-$$1)$$ \approx$$ 0.88$\,bit against $(5/3$$-$$\ln{3})
/\ln{2}$$ \approx$$ 0.82$\,bit for the balanced state.  Hence, when changing
from a discrete to a continuous field distribution, the metrological algorithm
profits from a refinement in the choice of initial states.  We note that any
initial state from $\mathcal{H}^3_{XY}$ yields the same (high) plateau level
but with the information gain for $t<T_s$ depending on the specific choice of
$\ket{\psi^0_{(1)}} \in \mathcal{H}^3_{XY}$; as these differences are small,
we do not pursue them further here.  In the second and further subsequent
steps, both scenarios maintain the feature of saturation in the information
gain per step, although some interesting new features appear, see Appendix \ref{fourier_cont} for a
detailed analysis of further optimization steps.

{\color{black} 
Ordinarily, it is the quantum Fisher information (QFI) estimate which is used as a performance metric. Here, we rather use the information gain as a performance metric which due to the postselection nature of our sensing protocol appears more appropriate.
As described above, the probability distribution of a magnetic field is updated according to the measurement conducted over different quantum states on each step of the algorithm.
While commonly used Cram\'{e}r--Rao bound, which is based on QFI, gives the asymptotic potential accuracy that can be achieved with a fixed quantum state, our protocol exploits a different approach and, therefore, can not be analyzed in the context of Fisher information in an evident way.
}

\section{Linear Ascending Metrological algorithm}
Expanding further on the above findings, we develop a novel quantum metrological
procedure exploiting phase coherence which ensures a near-Heisenberg limit
scaling even in the presence of dephasing processes when the Kitaev- and
standard Fourier protocols (see the descriptions in Appendixes \ref{kitaev} and \ref{fourier}) become ineffective.

The precision of a metrological algorithm is bounded by the maximum possible
number of iterations.  In the case of the standard Fourier- and Kitaev
procedures, the limitations are imposed by the coherence time $T_c$ and the {\color{black} characteristic}
duration of the control pulse $T_p$ {\color{black}(as described above, the evolution of the transmon qudit is determined by the applied rf-pulses, which cannot be made arbitrarily short and are defined by the hardware)} -- these times
bound the Ramsey delay on the longest and shortest time steps of the
procedure, respectively.  Since in both procedures the delay time of each
consequent step changes exponentially \textcolor{black}{(meaning that at each consequent step the delay time is either increased or decreased by factor $d$)}, the number of steps which can be
realized in practice is small \cite{danilin}.  One could then think of
performing multiple PER steps with the same delay time, as the recent
state-of-art techniques enable rather rapid transmon measurement- and
re-initialization procedures \cite{magnard, elo}.  However, such a routine is
essentially classical {\color{black} -- hereinafter, this routine will be referred to as the classical procedure --}
and its precision is therefore restricted by the shot
noise limit \cite{sekatski}. To overcome these problems, we propose a novel
Linear Ascending Metrological Algorithm or LAMA for qudit sensors that
combines both classical productivity and quantum scalability and, furthermore,
surpasses the efficiency of the Fourier- and Kitaev algorithms in common
realistic scenarios.  The $i$th step of the LAMA includes the following PER
sequence:
\begin{figure}
  \noindent\centering{
    \includegraphics[width=86mm]{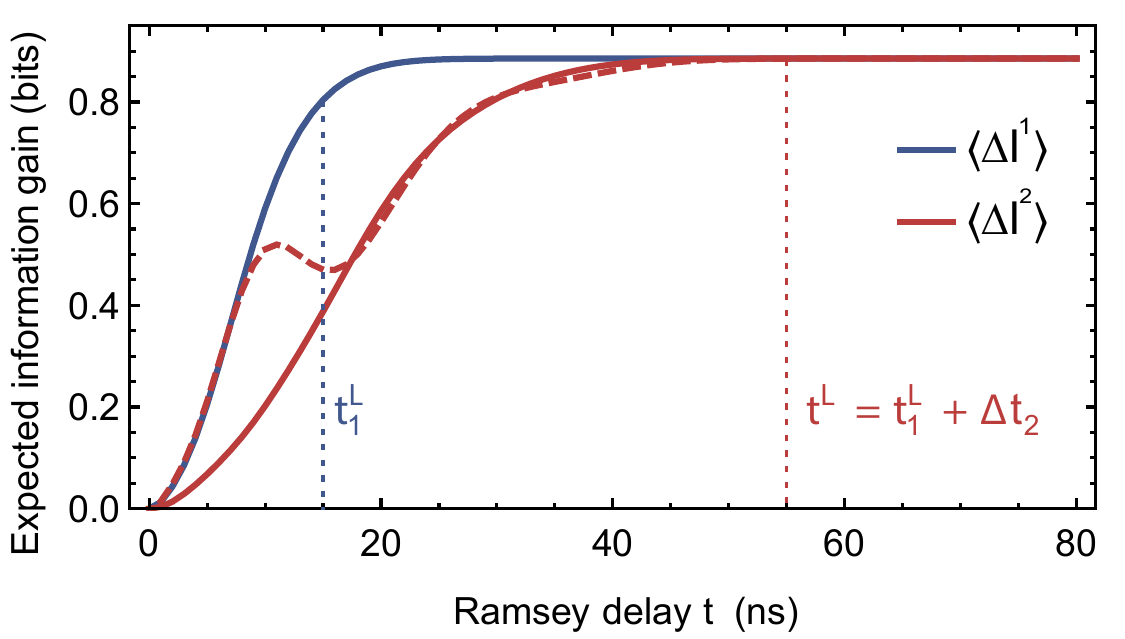}
    }
    \caption{First two steps of LAMA. Expected information gain
    $\langle \Delta I^{1,2} \rangle$ in the first (blue) and second (red)
    steps of the LAMA as a function of Ramsey delay time $t$.  The initial
    qutrit state is $\ket{\psi^0}=(1,\,\sqrt{2},\,1)^T/2\in
    \mathcal{H}^3_{XY}$ and the initial field distribution is a Gaussian
    $\mathcal{P}^{0}\big(\omega\,|\,\o\big) = N(0,\,\sigma^2)$ with
    $\sigma=2\pi/$($90$\,ns).
    $t^L_1$ and $t^L_2$ are the selected delay
    times on the first and second steps, respectively. The gain in the second
    step depends on the outcome of the first step, see dashed red line when
    $0$ is measured and solid red line for outcomes $1$ or $2$.}
    \label{new_alg_expected_gain}
\end{figure}

\begin{itemize}
    \item[\textbf{P}~] The qudit is prepared in an initial state
    $\ket{\psi^0}$ within $\mathcal{H}^d_{XY}$ (the same in every step). In the
    qutrit case, $\ket{\psi^0} = \frac{e^{i\alpha}}{2} (e^{i\beta}\ket{0} +
    \sqrt{2}\ket{1}+e^{-i\beta}\ket{2})$, where $\alpha$ and $\beta$ are real
    numbers.
    
    \item[\textbf{E}~] The qudit is exposed to the magnetic field during the
    predefined (rather than optimized) time interval $t_i^L = t_{1}^L +
    (i-1)\,\Delta t$, 
    {\color{black} with $\Delta t$ chosen of order $\sim$ $1/ \delta\omega_0$, where $\delta\omega_0$ is the initial field-uncertainty and $t_{1}^L=\max{(T_p, T_s)}$.}
    
    \item[\textbf{R}~] The qudit is subjected to the Fourier transform
    $\hat{F}_{d}$, 
    \begin{equation}
        \hat{F}_{d}|n\rangle=\frac{1}{\sqrt{d}} \sum_{k=0}^{d-1}
    \mathrm{e}^{-2 \pi i n k / d}|k\rangle,
    \end{equation}
    and a subsequent single-shot
    projective measurement of its state in the computational basis.  The
    probability distribution of the magnetic field is updated in accordance
    with Bayes formula.
\end{itemize}
A great simplification of the LAMA is the use of a fixed parameter set ${\bf
s}_i = {\bf s}$ for preparation and readout that does no longer require
further optimization after each step.  The following discussion of the
algorithm concerns again the case of a {\it continuously} field and refers to
the example of a qutrit with the optimized initial state $\ket{\psi^0} =
(1,\,\sqrt{2},\,1)^T/2\in \mathcal{H}^3_{XY}$ ($\alpha=\beta=0$) different
from the balanced one. While the qubit implementation of the new algorithm
cannot make use of a further optimisation of the initial state, the linear
increase in the delay time $t_i$ characteristic of the LAMA is still
advantageous.

\begin{figure*}[t]
  \noindent\centering{
    \includegraphics[width=170mm]{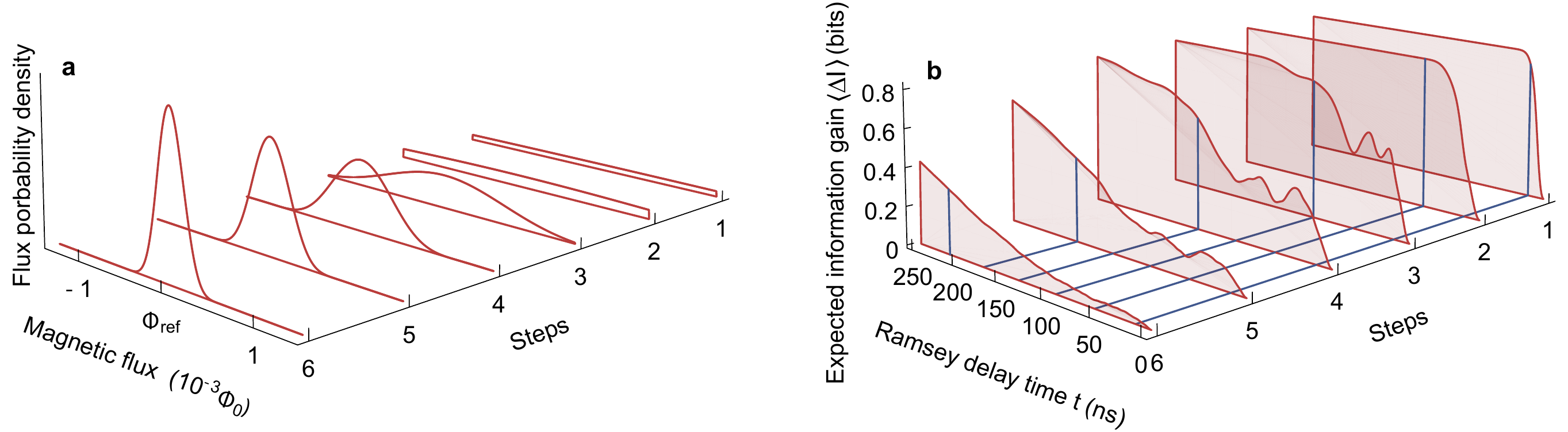} }
    \caption{LAMA operation. (a) field distributions and
    (b) expected information gains (before subsequent measurement) at
    six different steps of the new algorithm as obtained through numerical
    simulation without dephasing. The initial distribution function is
    $\mathcal{P}^{0}\big(\omega\,|\,\o\big)=N(0,\,\sigma^2)$ with
    $\sigma=2\pi/$($90$\,ns); the outcomes of the steps are chosen
    $\{\xi_1,\,\xi_2,\,\dots,\,\xi_6\}=\{0,\,0,\,\dots\,0\}$.  Given our
    choice of small delay times, the field distribution in all of the steps is
    given by an ever-narrowing single peak.  We translate the value of
    reduced magnetic field $\omega$ into the magnetic flux
    $\Phi = \omega_{01}\,|d\Phi/d\omega_{01}|$ ($\omega_{01}$ is
    the transition frequency) by putting \cite{danilin} $\mu=10^5\,\mu_0$,
    where $\mu_0$ is Bohr magneton.  In (b), the delay times at each
    step are marked by the vertical lines.  }
    \label{3d}
\end{figure*}

To see how the algorithm works, we first examine the case of an ideal hardware
in the absence of decoherence, $T_c\rightarrow\infty$.
Fig.\,\ref{new_alg_expected_gain} compares
the information gain in the second step (red; solid and dashed lines
correspond to different outcomes of the first step) with the result obtained
in the first step (blue) of the algorithm as function of the delay time $t$.
In order to obtain the maximal information in the first step within the
shortest time interval, we choose the delay time at the onset of the
saturation plateau, {\color{black} $t_1^L = T_s$.}
{\color{black} 
The saturation time $T_s \sim 1/\sigma$ is determined by the expected search range of magnetic fields $[\omega_{min}, \omega_{max}]$ that represents the dynamic range of the sensing device.
In Fig.\,\ref{new_alg_expected_gain}, as well as in the following numerical analysis, we set $\sigma=2\pi/$($90$\,ns) that ensures the maximum achievable dynamic range at reasonable saturation time of the expected information gain.
This choice of $\sigma$ gives $T_s\approx15$\,ns which corresponds to the fastest possible manipulations of transmons -- it becomes unfeasible to control superconducting devices using $T_p < 10^{-8}$\,s rf-pulses, thus, dynamics at shorter times is not meaningful for metrological purposes.
}
In the second step, the expected information gain again saturates at large delay
$t^L_2$, but is reduced at small times $\sim t^L_1$.  This reduction expresses
the fact that we have already obtained some information associated with the
previous delay time $t^L_1$; nevertheless, due to the probabilistic nature of
the quantum procedure, the prospective information gain remains non-zero.
Depending on the outcome of the first step, the gain in the second step may or may
not exhibit an additional peak at a delay time below $t^L_1$, see dashed line.
The appearance of such a new peak in the information gain in step $2$ depends
on the relation between the previous outcome $\xi_1$ and the chosen initial
state for step 2 within the XY-plane.
\textcolor{black}{
The nature of such dependence is the following: we update the field probability distribution based on the measurement outcome, reducing the distribution’s variance at each step. 
The mean of the distribution indicates the most probable magnetic field value and it drifts to the true value during the sensing procedure. 
For the qutrit case, while $\xi_1=0$ outcome does not change the mean of the distribution, $\xi_1=1$ and 2 outcomes shift the mean on the right and on the left respectively.
Besides this feature, with a proper choice of ${\bf s}_2$ the additional peak in the information gain appears at $3t_1^L$ ($dt_1^L$ for a qudit), where the next step of the Kitaev procedure should be conducted; the correspondence of the appearing peak with
the standard Fourier-based procedure is discussed in the \ref{fourier_cont}.
}
Instead of attempting to extract this peak information gain by choosing the new initial
state parameters ${\bf s}_2$ in accordance with the previous outcome, 
{\color{black} we prepare the qutrit in the same initial state (with parameters ${\bf s}_1$), and
adopt a time step $t^L_2=t^L_1+\Delta t$ with $\Delta t = C/\delta\omega_0$, $\delta\omega_0 = \sigma$.
Evaluation of the optimal prefactor $C$ for each step requires substantial computational time, therefore we will not concentrate on this optimisation. Instead, we numerically find the optimal $C\approx\pi$ and fix it for the whole metrological procedure.
As we will show further, this  choice of a constant time step already ensures the performance which, in the presence of decoherence, beats both classical and Fourier/Kitaev quantum procedures.}
This choice of $\Delta t$ and $t^L_2$ again
enables us to exploit the information gain near the plateau of step 2 and
learn nearly $0.88$ bit of information. The subsequent steps follow the same
route: regardless of the previous outcomes, the qutrit is always prepared in
the same state parametrized by ${\bf s}_1$, while the linear increase of the
delay time allows to operate away from the emerging drops in the information
gain, ensuring the advantage of the LAMA over the classical procedure. This
straightforward algorithm then provides a great simplification as compared to
the other algorithms involving Bayesian learning.

Shown in Fig.\,\ref{3d}(a) are the numerically simulated probability
distributions for the magnetic field for a six-step procedure; in turn,
Fig.\,\ref{3d}(b) displays a series of expected information gains, see Eq.\
\eqref{AvGain}, before the next measurement.  As the algorithm proceeds, the
extracted information per step decreases below the saturation limit of 0.88
bit with the plateau level first shifting further out to longer delay times and
then decreasing over the entire time interval.  Nevertheless, as shown below,
the practical realization of the LAMA can be quite beneficial in terms of the
total information accumulation and scalability.  More 3D plots such as
Fig.\,\ref{3d} for different sets of outcomes are presented in Appendix \ref{lama_op}.

\begin{figure*}[t]
  \noindent\centering{
    \includegraphics[width=85mm]{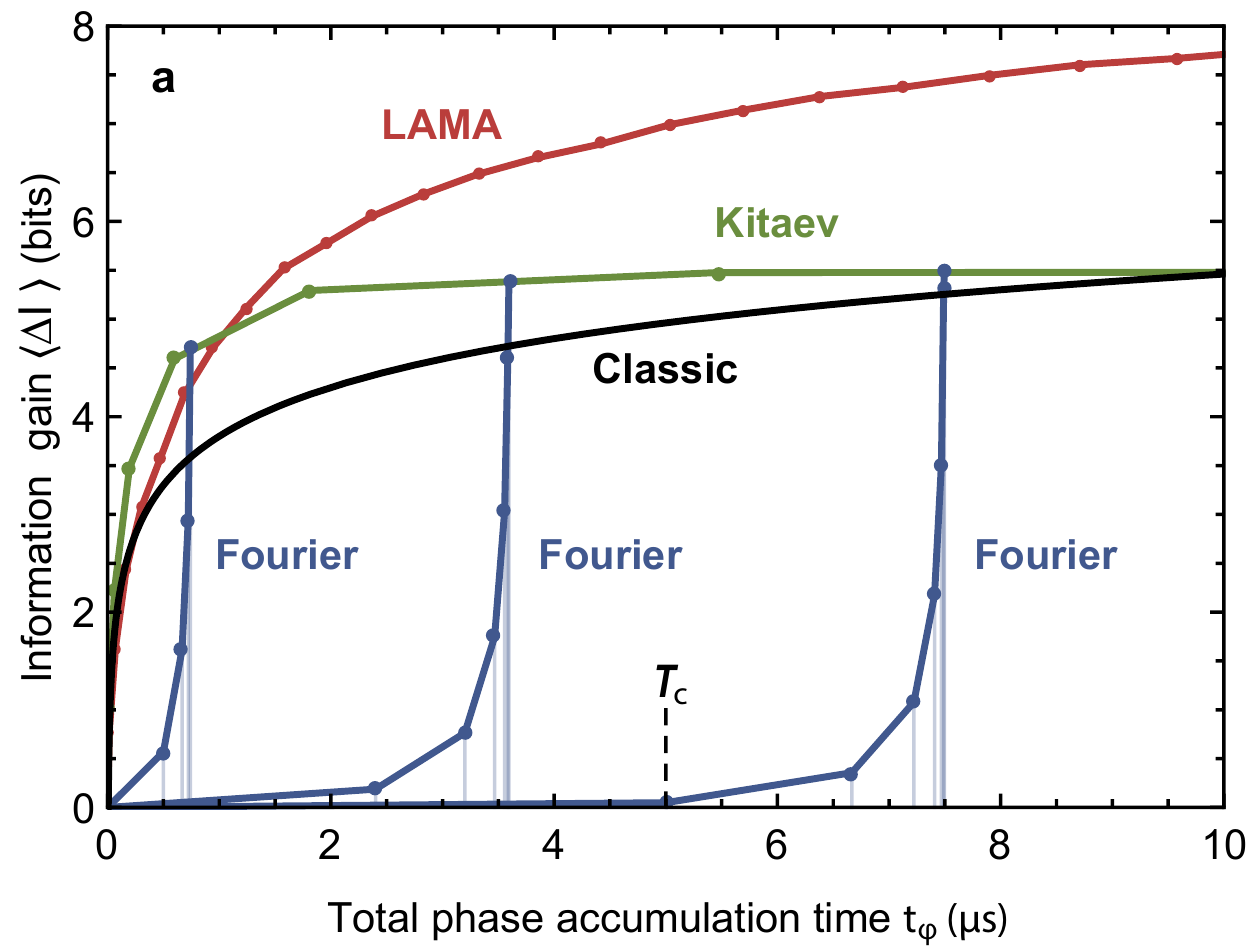} \quad
    \includegraphics[width=85mm]{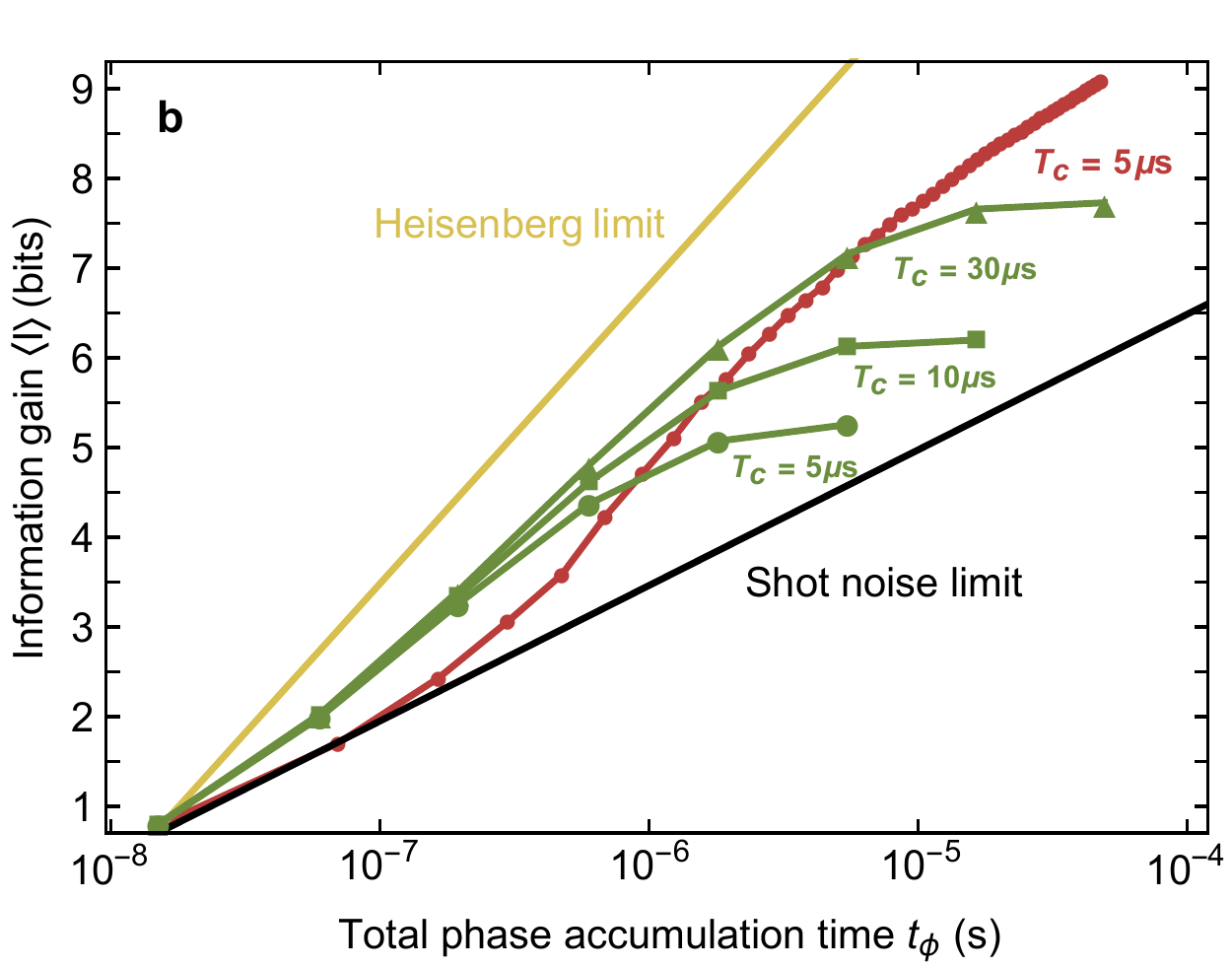}
    }
    \caption{Comparison of the algorithms {\color{black}for a transmon qutrit}. (a) Linear and (b)
    Linear-log plots of the information gain
    {\color{black} defined in Eq.\,\ref{AvGain}     
    in the classical (black), Kitaev
    (green), and standard Fourier (blue) procedures as functions of the total phase accumulation time;
    the red curve provides the result for the new LAMA.  
    }
    The symbols mark the
    individual steps of the four procedures.  All curves are averaged over a
    series of $10^3$ numerical experiments (for details of the simulation see
    the main text). We have chosen $T_s = 15$\,ns, $\mathcal{P}^{0}
    \big(\omega\,|\,\o\big) = N(0,\,\sigma^2)$ with $\sigma = 2\pi/$($90$\,ns)
    and $T_p < T_s$. 
    {\color{black} The initial delay time in the classical, $t^C_1$, and Kitaev, $t^K_1$,
    procedures  and in the LAMA, $t^L_1$, is $15$\,ns; in
    the case of the LAMA, the time increase per step is $\Delta
    t = 40\,$ns.  }
    In (a) the standard Fourier
    procedure is represented by three curves corresponding to different
    initial delay times: $t_1^{F} = 0.5$\,$\mu$s, $2.4$\,$\mu$s, and
    $5$\,$\mu$s.
    {\color{black} In the case of the Fourier procedure, a larger $t_1^{F}$ allows to perform more steps of the
    algorithm since on each step we are decreasing time from $t_1^{F}$ to $T_s$. However, due to the decoherence, it also results in an insufficient information gain in the first steps, $t_1^{F}$ approaches $T_c$. 
    In the case of the Kitaev procedure, decoherence affects last steps since they are performed at larger delay times. 
    The final gain in the Fourier case corresponds to the gain obtained with the Kitaev procedure at the same total phase accumulation time. Note that the highest points of blue curves collapse onto the green curve.  This is not a surprise since the Fourier and Kitaev procedures are expected to provide the same performance; for that reason we do not plot curves for the Fourier procedure in (b).}
    All curves in (a) are obtained for the same coherence time $T_c =
    5\,\mu\text{s}$, whereas in (b), we show three green curves 
    corresponding to the Kitaev procedure for three different values of $T_c$,
    5\,$\mu$s (circles), 10\,$\mu$s (squares), and 30\,$\mu$s (triangles); the
    red and black curves referring to the LAMA and to the classical procedure are
    obtained for $T_c = 5\,\mu\text{s}$.  The yellow straight line marks the
    Heisenberg quantum limit with scaling parameter $\alpha$ = 1.
   }
    \label{gain_vs_acum_time}
\end{figure*}

Next, we compare the LAMA with the \textcolor{black}{existing} metrological procedures for the
experimental situation where the initial field distribution is continuous and
the qutrit is subject to decoherence, i.e., $T_c$ is finite; details on how
our model accounts for dephasing and relaxation are given in Appendix \ref{dephasing}.  For simplicity, we assume that the minimal pulse duration $T_p$ is
much smaller than the saturation time $T_s$ in the first step, what limits the
delay time $t^L_1=\max{(T_p,\,T_s)}$. We base our comparison on a numerical
simulation constituting a series of $10^3$ separate computational experiments.
In each experiment, we numerically perform 50 sequential steps of the LAMA in
accordance with the above scheme, keeping the parameters $T_c$, $T_s$,
$t^L_1$, and $\Delta t$ the same for every experiment.  The outcomes of the
simulated experiments are generated randomly: in the $n$th step, the computer
samples the outcome $\xi_n$ from the probability distribution
$P(\xi_n\,|\,t_n) = \int P(\xi_n\,|\,\omega^\prime,\,t_n) \,\mathcal{P}^{n-1}
\big(\omega^\prime\, |\,\{\xi_i,t_i\}_{i=1}^{n-1}\big) \, d\omega^\prime$.  To
determine the efficiency of the procedure, we compose the results of all
computational experiments, i.e., we plot the information gain averaged over
the series of individual experiments as a function of the total phase
accumulation time.  We compare the LAMA with other algorithms through
simulating the operations of the latter in an analogous manner, although the
number of steps in the individual experiments may change as required by the
different algorithms.

The results of the simulations are summarized in
Fig.\,\ref{gain_vs_acum_time}(a), where we show the total information gain as a
function of the phase accumulation time $t_\phi =\sum_i t_i$ (the
sequence $t_i$ denotes all Ramsey delay times).  The plot
demonstrates that for a typical device and almost any total phase
accumulation time $t_\phi$, the LAMA allows for a larger information gain, thus
providing a higher precision, as compared with the more intricate standard
Fourier- and Kitaev protocols.  This is particularly evident for large
accumulation times $t_\phi \gtrsim T_c$ when the LAMA surpasses other algorithms
by $\gtrsim 30\%$, whereas the standard Fourier- and Kitaev procedures provide no
significant advantage over the trivial classical procedure.

Finally, we analyse the scaling behaviour of the new LAMA.  The efficiency of
a metrological procedure is reflected in the dependence of the
field-uncertainty  upon the total phase accumulation time \cite{danilin},
$\delta\omega(t_\phi) \propto t_\phi^{-\alpha}$.  A proper quantum algorithm
should exceed the shot noise limit of $\alpha=0.5$, ideally reaching the
Heisenberg limit $\alpha=1$.  Since for any accumulation time $t_\phi$ the
total information gain is given by $I \simeq -\ln{[\delta\omega(t_\phi)]} +
\ln{[\delta\omega(0)]}$, see Eq.\,(\ref{H}), the scaling can be analyzed in a
plot of $I$ as function of $\ln{t_\phi}$.  The lin-log plot in
Fig.\,\ref{gain_vs_acum_time}(b) displays the scaling behaviour of the classical
(in black, same curve for any (classical) delay time $t^C \ll T_c$; if $t^C
\gtrsim T_c$, the procedure operates below the shot noise limit) and standard
Fourier/Kitaev procedures (in green, three curves for
$T_c=\{5,\,10,\,30\}\,\mu s$), as well as the new LAMA (in red, $T_c = 5\,\mu
s$) for the continuous field with decoherence included.  The curves
corresponding to the standard Fourier- and Kitaev procedures (green) are
indistinguishable. The scaling parameter $\alpha$ can be obtained from the
slope of a tangent to a curve at any particular point.  As one expects, for
any $T_c$ larger than the delay time of a step, the classical procedure
complies with the shot noise limit, $\alpha=0.5$.  We also see that,
although for $t_\phi \ll T_c$ the standard Fourier and Kitaev procedures
approach the Heisenberg limit, for $t_\phi \sim T_c$, the value of $\alpha$
drops almost to zero. In turn, the new LAMA, though behaving classically for
small $t_\phi$, becomes much more efficient than the Fourier/Kitaev procedures
for $t_\phi \sim T_c$, with $\alpha$ reaching its maximum $\simeq 1$ in an
intermediate region at $t_\phi \simeq 1.1\, T_c$, at which point the delay time
per step $t^L_i\simeq T_c/7$.  One can see that a finite coherence time $T_c$
does not posit any notable limitation on the LAMA's efficiency until the delay
time $t^L_i$ becomes comparable with $T_c$; at large accumulation $t_\phi \sim
10\, T_c$ the algorithm still operates close to the shot noise limit with
$\alpha \gtrsim 0.5$.  Importantly, we see that even when implemented on a
mediocre transmon qutrit device with $T_c = 5$\,$\mu$s, the new LAMA is
capable to outperform other algorithms realized with a cutting edge device
with $T_c = 30$\,$\mu$s \cite{peterer}.

\section{Conclusion}
In summary, we have proposed a simple and robust sequential quantum
metrological algorithm for magnetometry, the LAMA, which is characterised by
two important features, (i) each measurement step involves a linear increase
of the Ramsey delay time, and (ii) the qudit is always initialized in the same
state of maximal spin perpendicular to the field. The prescription of these
measurement parameters drastically reduces the complexity of the algorithm as
compared to algorithms requiring learning. The linear increase in the Ramsey
delay time guarantees an improved performance as compared to the classical
algorithm, where the repeated measurement at the same Ramsey time produces a
steadily reduced information gain with each additional step. 
We have compared our algorithm with quantum Fourier and Kitaev algorithms and
demonstrated that the LAMA provides a markedly better performance in the
realistic situation where the qudit is subject to decoherence and the measured
field is distributed continuously. 
\textcolor{black}{As the decoherence limits the delay time range, the LAMA should enable us to utilize the resource of quantum coherence more effectively: by comparison with other algorithms, the LAMA better spans the full available range of times and thus allows for more iterations and better scaling. The presented results have far reaching implications going beyond the context in which they were derived.}
We anticipate our findings to accelerate the
progress towards reliable quantum magnetic sensors and find use in other
applications.

\section*{Acknowledgments}

We are grateful to S.\,N. Filippov, G.\,G. Amosov, and G.\,S. Paraoanu for valuable discussions.
This work was supported by the Government of the Russian Federation
(Agreement 05.Y09.21.0018), 
17-02-00396A, 18-02-00642A and 19-32-80005 (N.S.K. and M.R.P.),
Foundation for the Advancement of Theoretical Physics and Mathematics "BASIS",
the Ministry of Education and Science of the Russian Federation
16.7162.2017/8.9.
This work was also supported by the U.S. Department of Energy, Office of 
Science, Basic Energy Sciences, Materials Sciences and Engineering Division 
(V.M.V.) and the Swiss National Foundation via the National Centre of 
Competence in Research in Quantum Science and Technology (NCCR QSIT), 
and the Pauli Center for Theoretical Physics (G.B.).

\section*{Contributions}
M.R.P. and N.S.K. contributed equally to this work.
N.S.K. proposed the algorithm.
M.R.P., N.S.K., and V.V.Z. performed calculations and analyzed the numerical data.
N.S.K., M.R.P., V.M.V., and G.B. wrote the manuscript with inputs from all authors.
G.B.L., G.B., and V.M.V. supervised the project.
All authors discussed the results and contributed to the work.


\appendix

\section{Transmon phase accumulation dynamics}
\label{transmon}

Here, we discuss the working principles of the transmon device in the
context of magnetic field sensing;  an extended discussion on the subject can
be found in  \cite{shlyakhov}.  Let us examine the evolution of
the transmon device operating in a qudit mode (having effectively $d$ energy
levels) in the presence of the magnetic field.  The transmon's transition
energies can be described to leading order by the expression
\begin{multline}
    E_{n,n+1}(\omega) = E_{n+1}(\omega) - E_n(\omega)
    \\= \sqrt{8\,E_J(\omega)\,E_c} - E_c\,(n+1),
\end{multline}
where $E_n$ is the energy of the $n$th level, $E_J(\omega)$ is the Josephson
energy sensitive to the field $\omega$, and $E_c$ is the charging energy.
Note, that the dependence of the different transition energies on the magnetic
field is identical, as it is determined solely by $E_J(\omega)$.
%
%
As a result, the first-order correction to the energy separation $E_{0,n} =
E_n-E_0$ in the presence of the magnetic field is given by
\begin{equation*}
    \partial_\omega E_{0,n}\,\omega = n\, \partial_\omega E_{0,1}\,\omega
\end{equation*}
for any $n\in\{0,\,\dots,\,d-1\}$.  Thus, the phase accumulation due to the
field is linear in $n$ and we obtain the following expression for the effective
evolution operator in the rotating frame approximation,
\begin{equation}
\hat{U}^e=e^{-i \left(\mathrm{diag}\left[0,\,\partial_\omega E_{0,1}\omega t,\,2\,\partial_\omega E_{0,1}\omega t,\,\dots,\,(d-1)\,\partial_\omega E_{0,1}\omega t\right]\right)}.
\end{equation}
For simplicity, in what follows we put $\partial_\omega E_{0,1}=1$.  The above
linear phase-accumulation dynamics is crucial for the implementation of all
algorithms discussed in the paper.  At the same time, the LAMA can be realized
on any other multilevel system beyond the transmon, provided that the level
separations scale equally in the magnetic field.
\bigskip

\section{Optimization method}
\label{optimization}

Here, we elaborate on the general qutrit realization of the PER protocol in
the $i$th step \cite{shlyakhov} (all matrices are represented in the
computational basis $\ket{k}$, $k=\{0,\,1,\,2\}$):
\begin{itemize}
    \item[\textbf{P}~] The qutrit is initialized through the application of a
    two-tone rectangular rf-pulse \cite{shlyakhov} to the ground state
    $\ket{0}$.  The unitary evolution induced by the pulse has the form
    \begin{equation}
    \label{pulse}
        \hat{U}^p_{(i)}=\mbox{Exp}\left(-i \left[
            \begin{array}{ccc}
            0 & \Delta_{1(i)}^p & 0 \\
            \Delta_{1(i)}^p & 2 \epsilon_{(i)}^p  & \Delta_{2(i)}^p \\
            0 & \Delta_{2(i)}^p & 0 \\
            \end{array}
        \right]\right),
    \end{equation}
    and results in the state $\ket{\psi^{0}_{(i)}}=\hat{U}^p_{(i)}\ket{0}$.
    Here, $\Delta_{1,2(i)}^p$ are effective transition amplitudes between
    qubit states and $\epsilon_{(i)}^p$ is related to the pulse frequencies;
    their optimal values can be found by numerical optimization \cite{shlyakhov}.
%
    
    \item[\textbf{E}~] The qutrit interacts with the external magnetic field
    $H$ during time $t_i$: $\ket{\psi^{0}_{(i)}} \to \ket{\psi^{t_i}_{(i)}}$.
    The field changes the phases inside the state vector: $\ket{k}\rightarrow
    e^{ik\omega t_i}\ket{k}$ ($k=\{0,\,1,\,2\}$), where we omitted a common
    phase factor.  In the absence of decoherence, the evolution operator is
    given by
    \begin{equation}
        \hat{U}^e_{(i)}=\mbox{Exp}\left(-i \left[
            \begin{array}{ccc}
             0 & 0 & 0 \\
            0 & \omega\,t_i  & 0 \\
            0 & 0 & 2\,\omega\,t_i \\
            \end{array}
        \right]\right).
    \end{equation}

    \item[\textbf{R}~] The qutrit undergoes a readout operation generated
    again by the pulse
    \begin{equation}
        \hat{U}^r_{(i)}=\mbox{Exp}\left(-i \left[
            \begin{array}{ccc}
            0 & \Delta_{1(i)}^r & 0 \\
            \Delta_{1(i)}^r & 2 \epsilon_{(i)}^r  & \Delta_{2(i)}^r \\
            0 & \Delta_{2(i)}^r & 0 \\
            \end{array}
        \right]\right).
    \end{equation}
    Finally, the information about the field value is extracted through the
    single-shot projective measurement of $\ket{\psi^{f}_{(i)}} =
    \hat{U}^r_{(i)} \ket{\psi^{t_i}_{(i)}}$ in the computational basis.
\end{itemize}
The above PER-sequence yields a fairly cumbersome expression for the final
state (before the measurement) $\ket{\psi^f_{(i)}} = \hat{U}^r_{(i)}
\hat{U}^e_{(i)} \hat{U}^p_{(i)} \ket{0}$ and we refrain from providing it
here.

The optimization of the $(n+1)$st step for a given delay time $t$ attempts to
find the preparation and readout parameters
$\textbf{s}_{n+1} = (\epsilon^p_{(n+1)}, \Delta_{1(n+1)}^p,
\Delta_{2(n+1)}^p, \epsilon^r_{(n+1)}, \Delta_{1(n+1)}^r,
\Delta_{2(n+1)}^r)$ that produce the maximum expected information gain on
that step,
\begin{multline}
   \langle \Delta I^{n+1}\big(\{\xi_i,t_i,\textbf{s}_i\}_{i=1}^{n}, 
   \{\tilde\xi_{n+1}, t,\,\textbf{s}_{n+1}\} \big) \rangle 
   =\\ \max_{\textbf{s}}\, \langle  \Delta I^{n+1}
   \big(\{\xi_i,t_i,\textbf{s}_i\}_{i=1}^{n}, 
   \{\tilde\xi_{n+1}, t, \textbf{s}\} \big) \rangle,
\end{multline}
where the averaging is done in accordance with Eqs.\,(\ref{AvGain}) and
(\ref{AvEnt}).

To implement and optimize numerically the above procedure, we model the
continuous field by an evenly spaced grid with the large number of points
$M$$=$$10^5$; the initial field values are Gaussian weighted,
$\mathcal{P}^{0}\big(\omega\,|\,\o\big)$$=$$N(0,\,\sigma^2)$ with
$\sigma$$=$$2\pi/$(90\,ns). 
{\color{black} We choose $\sigma$ in a way that the saturation time of the expected information gain is $T_s\approx15$\,ns -- it becomes impracticable to apply gates faster than $10^{-8}$\,s to transmon devices.}
In the first step, the optimal choice of
$\textbf{s}_{1}$ for a given $t$ corresponds to the maximum
of the averaged information gain $\langle  \Delta I^{1}\big(\tilde\xi_{1}, t,
\textbf{s} \big) \rangle$.  Since for the numerical calculations the
continuous distribution is modelled by a discrete one, the integral with
respect to $\omega^\prime$ has to be replaced by a sum
\begin{multline}
   \langle  \Delta I^{1}\big(\tilde\xi_{1}, t, \textbf{s} \big) \rangle  
   = -\sum_{m=1}^M \mathcal{P}^{0}\big(\omega_m\,|\,\o\big)
   \ln{\mathcal{P}^{0}\big(\omega_m\,|\,\o\big)}\\
   - \sum_{\tilde\xi_{1}=0}^2 \sum_{m=1}^M S^{1}\big(\tilde\xi_{1}, 
   t, \textbf{s} \big)\,P(\tilde\xi_{1}\,|\,\omega_m, t, \textbf{s})\,
   \mathcal{P}^{0}\big(\omega_m\,|\,\o\big).
\end{multline}
Here $P(\tilde\xi_{1}\,|\,\omega_m, t, \textbf{s})$ is given by
Eq.\,(\ref{outcome}); the expression for $S^{1}\big(\tilde\xi_{1}, t,
\textbf{s} \big)$ can be obtained from Eqs.\,(\ref{recurrent}) and (\ref{H}),
\begin{multline}
   S^{1}\big(\tilde\xi_{1}, t, \textbf{s} \big) 
   = -\sum_{m=1}^M \frac{\mathcal{P}^{0}\big(\omega_m\,|\,\o\big)\,
   P(\tilde\xi_{1}\,|\,\omega_m, t, \textbf{s})}{\sum_{m^\prime=1}^M 
   \mathcal{P}^{0}\big(\omega_{m^\prime}\,|\,\o\big)\,
   P(\tilde\xi_{1}\,|\,\omega_{m^\prime}, t, \textbf{s})}\\ 
   \times\ln{\frac{\mathcal{P}^{0}\big(\omega_m\,|\,\o\big)\,
   P(\tilde\xi_{1}\,|\,\omega_m, t, \textbf{s})}{\sum_{m^\prime=1}^M 
   \mathcal{P}^{0}\big(\omega_{m^\prime}\,|\,\o\big)\,
   P(\tilde\xi_{1}\,|\,\omega_{m^\prime}, t, \textbf{s})}}.
\end{multline}
%
%
%

\section{Relaxation and dephasing processes}
\label{dephasing}

We extent our model to account for the decoherence processes appearing in the
transmons.  When the Ramsey delay time $t_i$ becomes comparable to the
coherence time $T_c$, the state analysis requires solving a kinetic equation,
which we choose of the Lindblad form \cite{Breuer2007},
\begin{equation}
   \frac{d\hat{\rho}}{dt}=-i [\hat{\rho},\hat{H}_{int}]
   +\Gamma_{01}\hat{D}[\sigma_{01}]\hat{\rho}
   +\Gamma_{12}\hat{D}[\sigma_{12}]\hat{\rho},
\end{equation}
with $\hat{H}_{int}$ the interaction Hamiltonian 
\begin{equation*}
        \hat{H}_{int}=\hbar\,\left(
            \begin{array}{ccc}
             0 & 0 & 0 \\
            0 & \omega  & 0 \\
            0 & 0 & 2\,\omega \\
            \end{array}
        \right),
\end{equation*}
$\Gamma_{ij}$ are the energy relaxation rates, and the superoperator $\hat{D}$
describes the process of energy relaxation,
\begin{equation}
   \hat{D}[\sigma_{ij}]\hat{\rho} 
   = \hat{\sigma}_{ij}\hat{\rho}\hat{\sigma}_{ij}^\dagger 
   - \frac{1}{2}\{\hat{\sigma}_{ij}^\dagger\hat{\sigma}_{ij},\hat{\rho}\},
\end{equation}
with the Lindblad operators $\hat{\sigma}_{ij}=\ket{i}\bra{j}$ ($i,\,j =
\{0,\,1,\,2\}$). In addition, we consider fluctuations $\delta H$ of the 
field due to fluctuating electric currents or magnetic impurities, assuming
Gaussian noise parametrized by the dephasing rate $\Gamma_{\varphi}$,
\begin{equation}
   \langle \delta H(t) \delta H(t)'\rangle 
   =\left(\frac{d\omega}{dH}\right )^{-2}\Gamma_{\varphi}\delta(t-t').
\end{equation}

The above extensions are incorporated in our PER-procedure in the following
manner:
\begin{itemize}
    \item[\textbf{P}~] In the $i$th step, the qutrit is put in the initial state
    defined by the density matrix $\hat{\rho}_{(i)}^0 =
    \ket{\psi^0_{(i)}}\bra{\psi^0_{(i)}}$.
    
    \item[\textbf{E}~] The qutrit interacts with the external magnetic field
    $H$ with corresponding changes of the phases in the density matrix, which
    is furthermore affected by the decoherence processes,
    $\hat{\rho}_{(i)}^0\rightarrow \hat{\rho}_{(i)} (t_i)$.  The elements
    ${\rho}_{(i)}^{pq}(t_i)$ $(p,\,q \in \{0,\,1,\,2\})$ of the resulting
    density matrix can be expressed through the elements of $\hat{\pi}_{(i)}
    (t_i)$ describing the qutrit state after the interaction in the absence of
    decoherence,
\begin{align}
    &{\rho}_{(i)}^{22}(t_i)=\pi_{(i)}^{22}(t_i)\,e^{-\Gamma_{21}\,t_i}, \\
    &\begin{aligned}
    {\rho}_{(i)}^{11}(t_i)=&\pi_{(i)}^{11}(t_i)\,e^{-\Gamma_{10}\,t_i} \\
    + &\pi_{(i)}^{22}(t_i)\,\frac{\Gamma_{21}}{\Gamma_{21}-\Gamma_{10}}\,
    \left(e^{(\Gamma_{21}-\Gamma_{10})\,t_i}-1\right),
    \end{aligned} \\
    &{\rho}_{(i)}^{00}(t_i)=1-{\rho}_{(i)}^{11}(t_i)-{\rho}_{(i)}^{22}(t_i),\\
    &{\rho}_{(i)}^{01}(t_i)=\pi_{(i)}^{01}(t_i)\,e^{-\frac{(\Gamma_{10}+\Gamma_{\varphi})\,t_i}{2}},\\
    &{\rho}_{(i)}^{10}(t_i)=\pi_{(i)}^{10}(t_i)\,e^{-\frac{(\Gamma_{10}+\Gamma_{\varphi})\,t_i}{2}},\\
    &{\rho}_{(i)}^{02}(t_i)=\pi_{(i)}^{02}(t_i)\,e^{-\frac{(\Gamma_{21}+4\Gamma_{\varphi})\,t_i}{2}},\\
    &{\rho}_{(i)}^{20}(t_i)=\pi_{(i)}^{20}(t_i)\,e^{-\frac{(\Gamma_{21}+4\Gamma_{\varphi})\,t_i}{2}},\\
    &{\rho}_{(i)}^{12}(t_i)=\pi_{(i)}^{12}(t_i)\,e^{-\frac{(\Gamma_{21}+\Gamma_{01}
    +\Gamma_{\varphi})\,t_i}{2}},\\ 
    &{\rho}_{(i)}^{21}(t_i)
    =\pi_{(i)}^{21}(t_i)\,e^ {-\frac{(\Gamma_{21}+\Gamma_{01}+\Gamma_{\varphi})\,t_i}{2}}.
\end{align}
\item[\textbf{R}~] The qutrit undergoes the readout operation
$\hat{U}^r_{(i)}$ and the final state is subjected to the single-shot projective measurement.
\end{itemize}

As an example, we consider the first step of the standard Fourier procedure in
the presence of dephasing.  Using the above formulas and Eq.\,(\ref{outcome}),
we can write the probability of finding qutrit in state $\ket{\xi_1}$ in the form
\begin{align}
    P(\xi_1\,|\,\omega,t) =& |\bra{\xi_1}\,\hat{U}^r_{(1)}\,\hat{{\rho}}_{(1)}(t_i)\,
    \hat{U}^{r\,\dagger}_{(1)}\,\ket{\xi_1}|^2\notag \\
    = &\frac{1}{3} + \frac{2}{9}\,\mbox{cos} \big(\omega\,t-\frac{2\pi}{3}\xi_1 \big)\,
    e^{-\frac{\Gamma_{10} +\Gamma_{\varphi}}{2}t}\notag\\ 
    + &\frac{2}{9}\,\mbox{cos} \big(2\omega\,t+\frac{2\pi}{3} \xi_1 \big)\,
    e^{-\frac{\Gamma_{21}+ 4\Gamma_{\varphi}}{2}t} \notag 
    \\+ &\frac{2}{9}\,\mbox{cos}\big(\omega\,t-\frac{2\pi}{3} \xi_1 \big)\,
    e^{-\frac{\Gamma_{10}+\Gamma_{12}+\Gamma_{\varphi}}{2}t}.
\end{align}
In our simulations, we assume $\Gamma_{01} = \Gamma_{12}/\sqrt{2} =
\Gamma_{\varphi} = \Gamma$ and describe the decoherence rate by the coherence
time $T_c = 1/\Gamma$.

\section{Fourier-based metrological algorithm}
\label{fourier}

The standard Fourier-based algorithm is a sequence of PER steps performed with
different delay times.  Including learning, the algorithm involves a
conditional initial preparation at each step that depends on the previous
outcome. The $i$th step of the base-3 procedure can be described by the
following scheme:
\begin{itemize}
    \item[\textbf{P}~] the qutrit is prepared in the state
    $\ket{\psi^0_{(i)}}=\frac{1}{\sqrt{3}}(\ket{0}+e^{i\alpha_i}\ket{1}
    +e^{2i\alpha_i}\ket{2})$,
    where
    $\alpha_i = -\frac{2\pi}{3}\left(\frac{\xi_{i-1}}{3^1}+\frac{\xi_{i-2}}{3^{2}}
    +\cdots +\frac{\xi_1}{3^{i-1}}\right)$ and $\xi_j$ is the outcome of the $j$th step 
    (and $\alpha_1=0$),
    
    \item[\textbf{E}~] the system is exposed to the magnetic field during time
    $t_i^F=t_{1}^F/3^{i-1}$,
    
    \item[\textbf{R}~] the qutrit undergoes the Fourier transform given by Eq.\,(\ref{F}).
    The subsequent single-shot projective measurement of the final state in the
    computational basis provides the new information on the field.  The
    probability distribution is updated in accordance with Bayes' formula.
\end{itemize}

In order to illustrate the principle of the algorithm, we consider the
situation where the measured field can be expressed in a ternary
decomposition,
\begin{equation}
    \omega=\omega_0\,\left(\frac{r_{K}}{3^{0}}+\frac{r_{K-1}}{3^{1}}
    +\cdots+\frac{r_{1}}{3^{K-1}}\right),
\end{equation}
where trits $r_n$, $n\in \{1,\dots,K\}$ can assume values 0, 1, and 2.

In the first step, the qutrit is prepared in the balanced state
$\ket{\psi^0_{(1)}} = \frac{1}{\sqrt{3}}(\ket{0}+\ket{1}+\ket{2})$ and is
exposed to the field for time $t_1^F = 2\pi\cdot3^{K-2}/\omega_0$.  The qutrit
thus assumes the state $\ket{\psi^{t_1}_{(1)}} =i \frac{1}{\sqrt{3}}(\ket{0} +
e^{i\phi_1}\ket{1} + e^{2i\phi_1}\ket{2})$ (we omit the overall phase factor
$e^{-i\omega t_1}$) where $\phi_{1}=(2 \pi / 3) r_{1}$ is the field-dependent
phase which can be unambiguously determined through the application of the
Fourier transform and a projective measurement of the final state
$\hat{F}\ket{\psi^{t_1}_{(1)}}=\ket{r_1}$; the latter constitutes one of the
computational basis vectors $\ket{0}$, $\ket{1}$, or $\ket{2}$. The outcome of
the measurement is $\xi_1=r_1$.

Turning to the second step and accounting for the first outcome, the qutrit is
prepared in the modified balanced state $\ket{\psi^0_{(2)}} = \frac{1}
{\sqrt{3}}(\ket{0} + e^{i\alpha_2}\ket{1} + e^{2i\alpha_2}\ket{2})$ with
$\alpha_2 =-\frac{2\pi r_1}{9}$.  After an exposure time $t_2^F = t_1^F/3$,
the qutrit evolves to the state $\ket{\psi^{t_1}_{(1)}} = \frac{1} {\sqrt{3}}
(\ket{0} + e^{i(\alpha_2+\phi_2)}\ket{1} + e^{2i(\alpha_2+\phi_2)}\ket{2})$,
where $\phi_2$ is the field dependent phase and $\alpha_2+\phi_2 =
-\frac{2\pi}{9}r_1 + \frac{2\pi}{3}(r_2 + \frac{r_1}{3}) = \frac{2\pi}{3}r_2$.
The digit $r_2$ can then be found through proper readout and measurement as in
the previous step.  Similarly, the subsequent steps provide the further
digits.

\section{Kitaev algorithm}
\label{kitaev}

While the standard Fourier procedure allows to progressively learn the ternary
value of the field starting from the `smallest' digit $r_1$, the Kitaev
algorithm works in the reversed manner outputting the leading digit $r_K$
first.  In the $i$th step of the base-3 procedure
\begin{itemize}
    \item[\textbf{P}~] the qutrit is prepared in the state
    $\ket{\psi^0_{(i)}}=\frac{1}{\sqrt{3}}(\ket{0}+\ket{1}+\ket{2})$,
    
    \item[\textbf{E}~] the system is exposed to the magnetic field during the 
    time $t_i^K=t_{1}^K \cdot 3^{i-1}$,
    
    \item[\textbf{R}~] the qutrit undergoes a Fourier transform $\hat{F}_3$
    given by Eq.\,(\ref{F}).  The information about the field value is
    extracted through the single-shot projective measurement of the final state in the
    computational basis and the probability distribution is updated in
    accordance with the Bayes formula.
\end{itemize}
As distinct from the standard Fourier procedure, with each step the delay time
increases, ensuring that the field distribution is always represented by a
single peak.

\section{Oscillatory features in the information gain}
\label{oscillations_sec}
\begin{figure}[t]
  \noindent\centering{
    \includegraphics[width=\columnwidth]{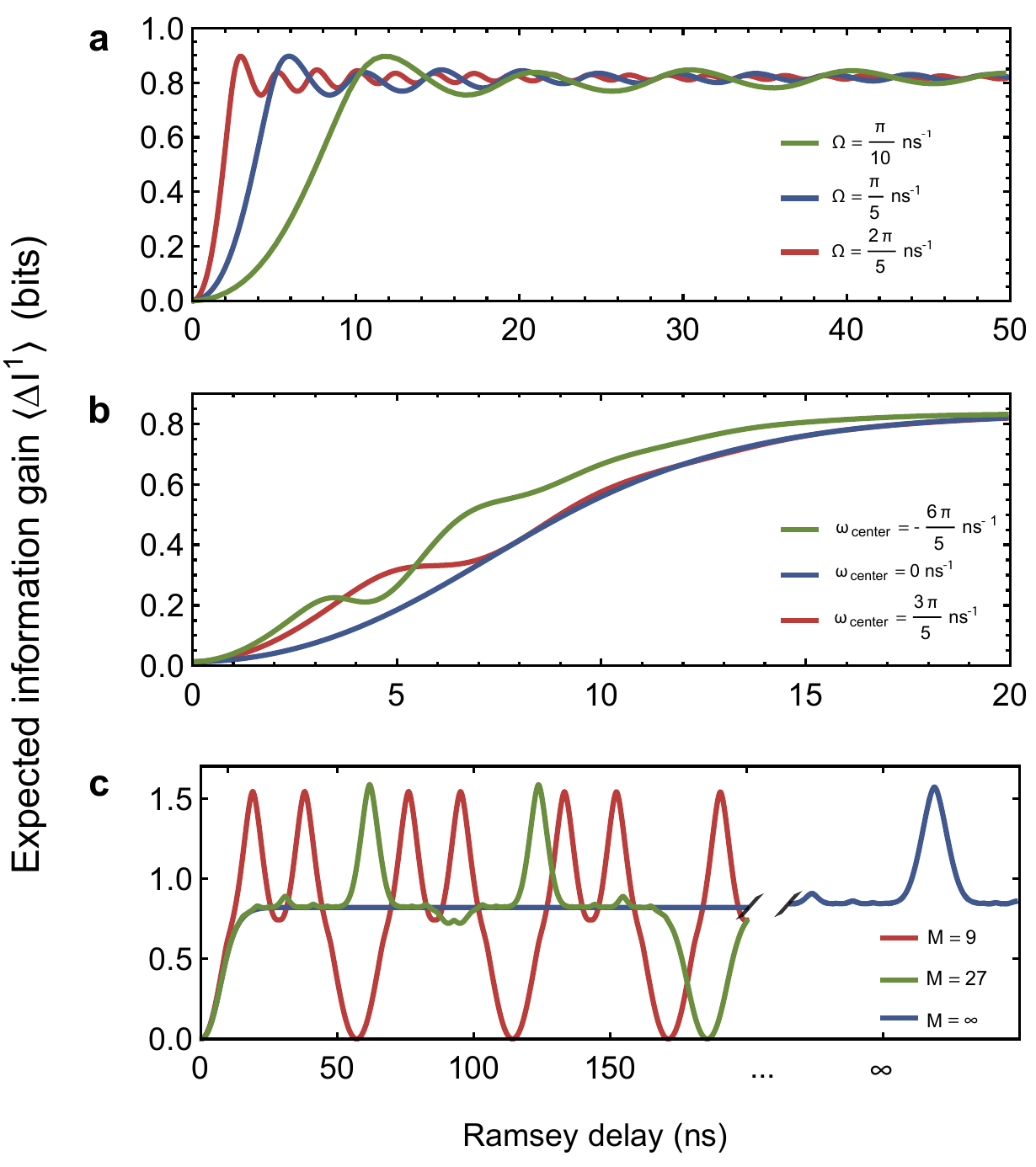}
    }
    \caption{Oscillations in the information gain. Expected
    information gain during the first step of the standard Fourier procedure
    as a function of the delay time evaluated under different conditions.  The
    detailed behaviour of the information gain function is determined by the
    initial field distribution.  (a) Information gain for a continuous
    uniformly distributed initial field with zero mean and continuous sampling
    for three different sharp intervals of width $\Omega$.  Oscillations with
    period $T_\mathrm{edge}\propto\Omega$ appear due to the sharp edges of the
    uniform distribution.  (b) Information gain for a continuous
    Gaussian initial distribution with $\sigma = 2\pi/$($90\,$ns) and
    continuous sampling for three different mean values
    $\omega_\mathrm{center}$.  The period of the observed oscillations is
    given by $T_\mathrm{center} \propto 1/|\omega_\mathrm{center}|$.
    (c) Information gain for a Gaussian initial distribution with
    $\sigma = 2\pi/$($90\,$ns) and zero mean and different sampling rates. The
    oscillatory pattern has a period $T_{\Delta}\propto 1/{\Delta\omega}$.  }
    \label{oscillations}
\end{figure}

In this section*, we discuss the oscillatory features intrinsic to the expected
information gain as function of the delay time.  The plots describing the
information gain and presented in the text (e.g., Fig.\,3(b)) have been
obtained under the assumption that initially, the field is distributed
continuously and the distribution function is Gaussian centered at $\omega=0$.
Under these conditions, no oscillatory features show up; the latter are not
central for the main discussion, but we shall address them now as they may
show up in other circumstances.  Given the complexity of the underlying
mathematical expressions, our numerical analysis has a rather qualitative
character.  The analysis concerns the base-3 standard Fourier procedure (with
no dephasing in the system), but the results are also relevant to the other
metrological algorithms proposed in the main text.

In general, there are three types of oscillations showing up in the
information gain related to the edges, the center position, and the
discretization scale of the distribution function.  We shall study each of them
separately by evaluating the information gain as a function of delay time
in the first step of the standard Fourier procedure under different
conditions.
\\
\textbf{Edges.--} 
We consider the situation where the field is distributed continuously over the
interval $[-\Omega/2,+\Omega/2]$.  Fig.\,\ref{oscillations}(a) displays the
information gain for a uniform distribution, with a saturation plateau
modulated by oscillations with period $T_\mathrm{edge}\propto 1/\Omega$. Such
oscillations are imposed by the abrupt edges of the distribution function and
disappear for the case of a Gaussian distribution with smooth tails.
\\
\textbf{Center position.--}
Another type of oscillations originates from a non-vanishing field average.
Fig.\,\ref{oscillations}(b), depicts the information gain for a continuous field
with a Gaussian distribution centered at different positions
$\omega_\mathrm{center}\neq 0$ and at $\omega_\mathrm{center} = 0$.  Again, oscillations with
a period $T_\mathrm{center}\propto 1/|\omega_\mathrm{center}|$ show up, this time in the
rising part of the gain function.
\\
\textbf{Discreteness.--}
The third type of oscillations appears for the case of a discretely
distributed field.  In our work, the continuous case is modelled with a
fine-grained discrete distribution built from outcomes with a large number of
measurable values.  Numerically, the continuous distribution on the interval
$[\omega_\mathrm{min},\,\omega_\mathrm{max}]$ implies that the field can
assume $M\sim 10^5$ possible values $\{\omega_\mathrm{min},
\,\omega_\mathrm{min}+\Delta \omega,\, \omega_\mathrm{min}+2\Delta
\omega,\,\dots,\,\omega_\mathrm{max}\}$ with $\Delta \omega =
(\omega_\mathrm{max}-\omega_\mathrm{min})/M$.  As revealed in
Fig.\,\ref{oscillations}(c), a non-zero spacing $\Delta \omega$ gives rise to
a periodic pattern with period $T_{\Delta}\propto 1/{\Delta\omega}$; the
depicted plots are obtained for the case of a Gaussian distribution on the
interval $[-\Omega/2,\,+\Omega/2]$ with different values of $M$.

The oscillations appearing in Figs.\,\ref{new_proc_3_outcomes}(b,\,d,\,f)
presented below are mainly of the second type, as after the first step, the
center of the distribution shifts away from zero.  In addition, oscillations
of the third type play an important role.  When the field values are discrete
with spacing $\Delta \omega > 0$, the standard Fourier procedure generates
peaks with the largest ones corresponding to the maximum possible information
gain of 1\,trit $=$  $\log_2(3)$\,bit.  This reaffirms that for the
discrete case, the standard Fourier procedure is indeed optimal, given an
appropriately chosen delay time.  However, this is no longer the case when the
field distribution is continuous and $\Delta \omega = 0$, since the
oscillatory peak would correspond to an infinite delay time.  Nevertheless, as
we will see in the next section, this type of oscillations still manifest
themselves in the second and subsequent steps of the procedure.

\section{Standard Fourier procedure in the continuous case}
\label{fourier_cont}
\begin{figure*}[t]
  \noindent\centering{
    \includegraphics[width=150mm]{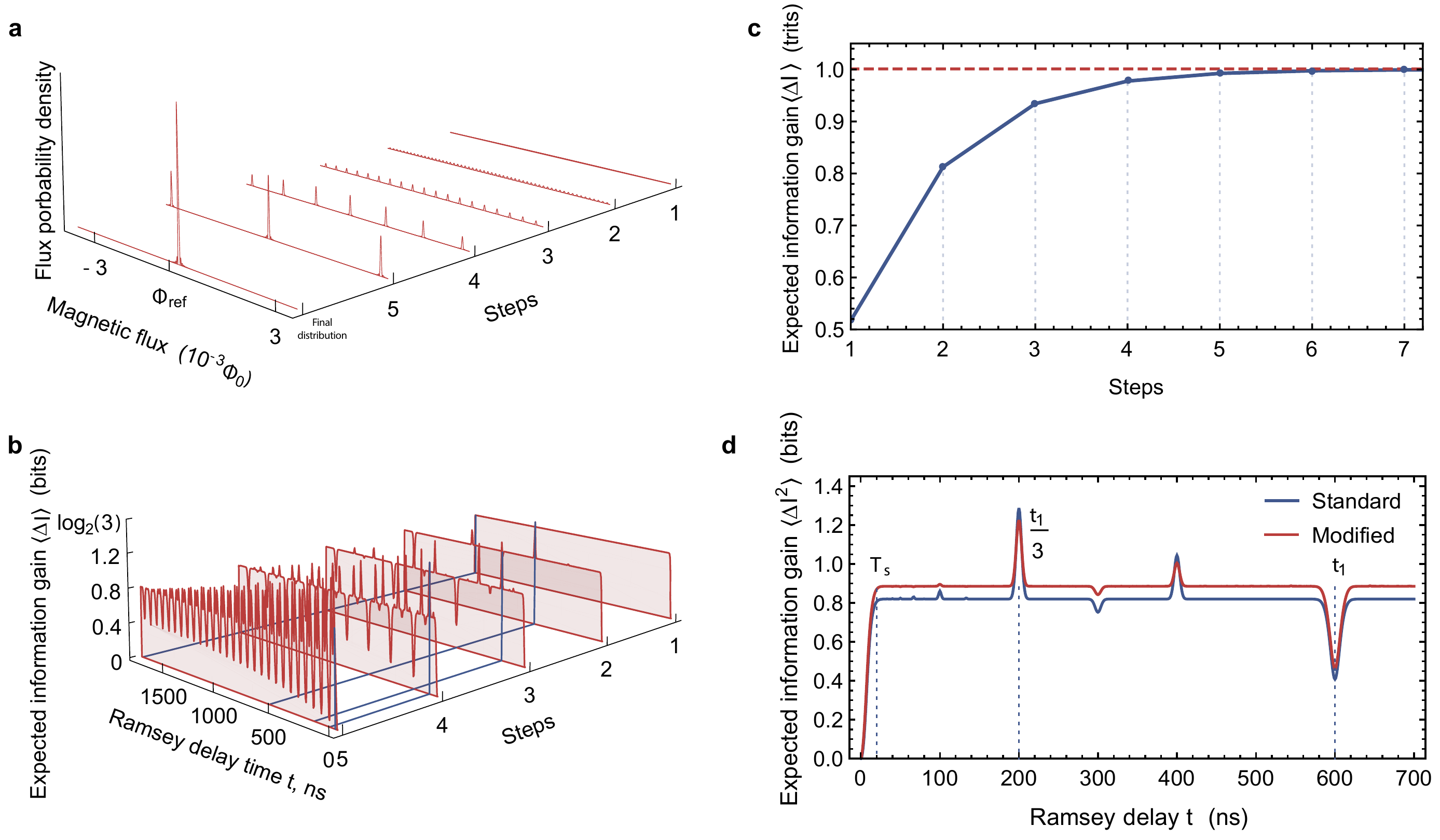}
    }
    \caption{Standard Fourier algorithm. (a) Field
    distributions and (b) expected information gains at six different
    steps of the standard Fourier procedure as obtained through numerical
    modelling.  In (b), the delay times at each step are marked by the
    vertical lines.  The initial distribution function is normal with
    $\mathcal{P}^{0}\big(\omega\,|\,\o\big)=N(0,\,\sigma^2)$ and a width
    $\sigma=2\pi/$($90$\,ns); dephasing is absent.  (c) The expected
    information gain per step (expressed in trits) in the standard Fourier
    procedure at each of the steps in the continuous case with no dephasing.
    The dashed red line indicates the maximum possible information gain of 1
    trit.  The information gain on the 7th step is 0.99 trits.  (d)
    Expected information gain in the second step of the standard (blue) and
    modified (with alternative preparation step, in red) Fourier procedures
    for a continuously distributed field with no dephasing processes.}
    \label{fourier_overall}
\end{figure*}

We now consider in more detail the second and further steps of the base-3
standard Fourier procedure for the case of a continuous field distribution
when the dephasing is absent.  With the delay time in the first step chosen
relatively large in comparison with the saturation time $T_s$, the field
distribution function becomes periodic and hence semi-discrete, see
Fig.\,\ref{fourier_overall}(a). These oscillations reflect in the behaviour of
the information gain of the second step, see Fig.\,\ref{fourier_overall}(b):
although it continues to be a saturation curve, it now exhibits oscillatory
peaks which we previously classified as of the third type; the largest of such
peaks corresponds exactly to delay time $t^F_2 = t^F_1/3$.  In the subsequent
steps, with the field distribution becoming increasingly more discrete-type,
the information gain per step tends asymptotically to the maximum of 1 trit,
see Fig.\,\ref{fourier_overall}(b,\,c).

We have previously learnt that in the first step, the balanced initial state
used in the standard Fourier procedure, is inferior as compared to chosing a 
state from $\mathcal{H}^3_{XY}$.  Given this fact, it is interesting to check
whether the standard Fourier procedure could be improved through a
modification of the preparation stage at each step.  We consider the
following modification of the standard procedure:

\begin{itemize}
    \item[\textbf{P}$^\prime$~] in the $i$th step, the qutrit is prepared in
    the state $\ket{\psi^0_{(i)}} = (\frac{1}{2}\ket{0} + \frac{e^{i\alpha_i}}
    {\sqrt{2}}\ket{1} + \frac{e^{2i\alpha_i}}{2}\ket{2}) \in
    \mathcal{H}^3_{XY}$, where $\alpha_i = -\frac{2\pi}{3} \left(
    \frac{\xi_{i-1}} {3^1} + \frac{\xi_{i-2}}{3^{2}} + \cdots + \frac{\xi_1}
    {3^{i-1}}\right)$ and $\xi_j$ is the outcome of the $j$th step
    ($\alpha_1=0$).
    
    \item[\textbf{E}~] the system is exposed to the magnetic field during time
    $t_i^F=t_{1}^F/3^{i-1}$.
    
    \item[\textbf{R}~] the qutrit is subjected to a Fourier transform
    $\hat{F}_3$ as given by Eq.\,(\ref{F}).  The information about the field
    value in extracted through the single-shot projective measurement of the final state
    in the computational basis.  The probability distribution is updated in
    accordance with Bayes' formula.
\end{itemize}
%
%
Fig.\,\ref{fourier_overall}(d) displays the information gain in the second step
for the two cases of standard (shown in blue) and modified (shown in red)
Fourier procedure.  It turns out that, although the modified algorithm
produces a higher saturation level and the conditional preparation allows to
extract information above the plateau, after the first step, the standard
procedure becomes more efficient as the peaks are larger in this case. One
should note, though, that this improved efficiency of the standard
algorithm can only be exploited with very precisely chosen delay times;
otherwise, if the time does not comply with the protocol, the information
would be extracted from the plateau-level, which is higher in the case of the
modified procedure.  Note also, that the outlined modified procedure is not
the same as the one proposed in the main text.

\section{LAMA operation}
\label{lama_op}

\begin{figure*}[t]
  \noindent\centering{
    \includegraphics[width=180mm]{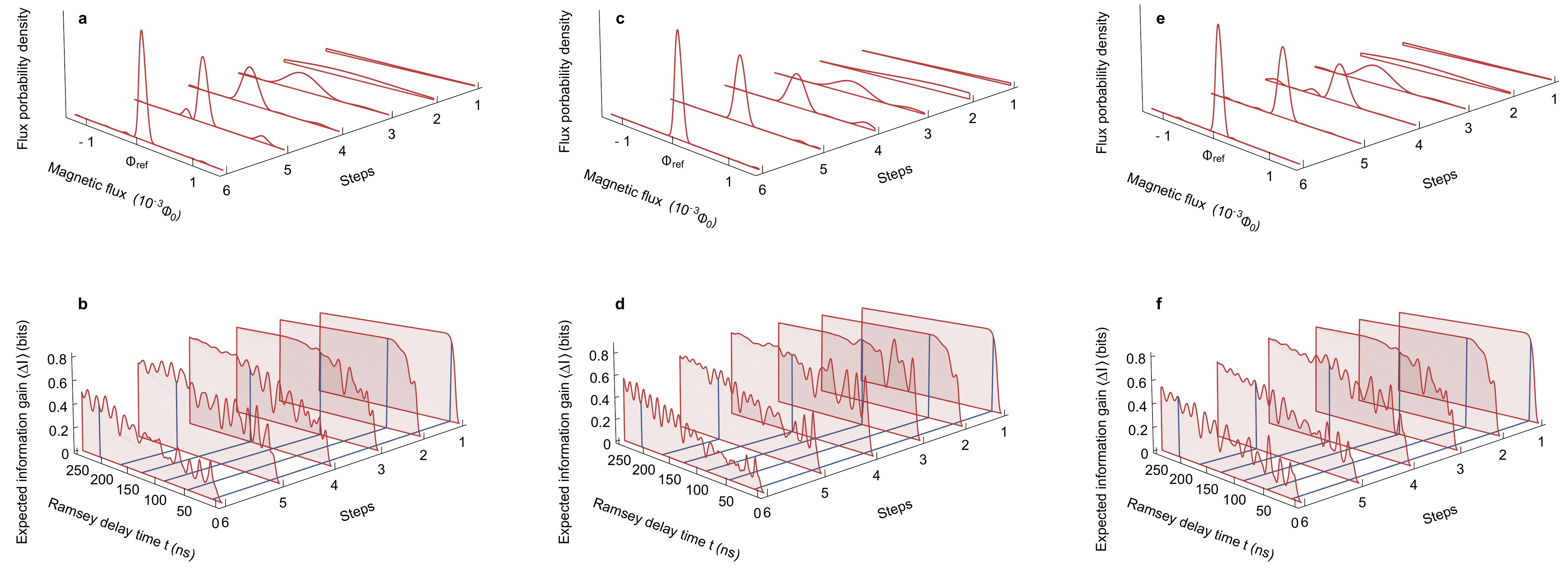}
    }
    \caption{LAMA operation. (a), (c), (e),
    Field distributions and (b), (d), (f), expected
    information gains at six different steps of LAMA obtained through
    numerical modelling in the absence of dephasing. The corresponding outcome
    sets ($\{\xi_1,\,\xi_2,\,\dots,\,\xi_6\}$) are (a), (b)
    \{0,2,0,1,1,2\}; (c), (d) \{1,1,1,1,1,1\}; (e),
    (f) \{1,2,0,0,1,2\}.  The initial distribution function is
    $\mathcal{P}^{0}\big(\omega\,|\,\o\big)=N(0,\,\sigma^2)$ with
    $\sigma=2\pi/$($90$\,ns).  } \label{new_proc_3_outcomes}
\end{figure*}

Shown in Fig.\,\ref{new_proc_3_outcomes}(a,\,c,\,e) are the numerically simulated
probability distributions for the magnetic field for different six-step
procedures; Fig.\,\ref{new_proc_3_outcomes}(b,\,d,\,f) display three corresponding series of expected
information gains before the next measurement.  The plots correspond to
different outcome sets, see caption, and illustrate the appearance of
oscillations in the second and latersteps due to the non-zero mean value of
the updated distribution.

%

%

\end{document}